\documentstyle[preprint,aps]{revtex}
\tighten
\begin{document}

\title{Flat foliations of spherically symmetric geometries}
\author{Jemal Guven${}^{(1)}$\thanks{e-mail \tt{ jemal@nuclecu.unam.mx}}
 and
   Niall \'O Murchadha${}^{(2)}$\thanks{e-mail \tt{ niall@ucc.ie}}}
\address{
${}^{(1)}$\
 Instituto de Ciencias Nucleares \\
 Universidad Nacional Aut\'onoma de M\'exico\\
 Apdo. Postal 70-543, 04510 M\'exico, D.F., MEXICO \\
${}^{(2)}$\ Physics Department,  University College Cork,\\
Cork, IRELAND \\
}
\maketitle
\begin{abstract}
We examine the solution of the constraints in spherically symmetric general
relativity when spacetime has a flat spatial hypersurface. We demonstrate
explicitly that given one flat slice, a foliation by flat
slices can be consistently evolved. We show that when the sources are
finite these slices do not admit singularities and we provide an explicit
bound on the maximum value assumed by the extrinsic curvature. If the dominant
energy condition is satisfied, the projection of the extrinsic
curvature orthogonal to the radial direction
possesses a definite sign. We provide both necessary and sufficient
conditions for the formation of apparent horizons in this gauge which are
qualitatively identical to those established earlier for extrinsic
time foliations of spacetime, {\it Phys. Rev. D} {\bf 56} 7658, 7666 (1997)
which suggests that these conditions possess a gauge invariant validity.
\end{abstract}
\date{\today}



\noindent{\bf  Introduction}
\vskip 0.5cm

Despite the impressive array of techniques we currently possess
for constructing initial data in general relativity, an
unsatisfying feature persists: the sensitivity of the
solution to the way we foliate spacetime. Indeed, this often
overshadows entirely the physics we are attempting to understand.
It is clear that the problem is even worse at the dynamical level.

In order to solve the constraints it is traditional to appeal to
an extrinsic time foliation of spacetime. This involves some
restriction on the extrinsic curvature so as to define the slicing.
One such foliation, in particular, has a long history: in
asymptotically flat spacetimes,  the mean extrinsic curvature is
set to zero. In the simplified context of spherically symmetric
general relativity it is possible to examine explicitly
how sensitively the solution of the
constraints depends on this choice of gauge.
This is easy in this case because the
extrinsic curvature is completely specified by two scalar functions
and the mean curvature is a linear combination of these two
scalars. One can then treat any extrinsic time
foliation as a specification of the ratio of these scalars.
This has been the approach adopted in \cite{I,II,III,IV,V}
in our examination of the constraints.

In \cite{III,IV,V} we restricted our attention to extrinsic
curvatures which,
at every point on the slice, are timelike vectors in
superspace. In these articles we were able to demonstrate that the
solution to the constraints is largely independent of the
particular extrinsic time foliation we choose. Remarkably, we
appear to have a robust characterization of the physical landmarks
of the spatial geometry; specifically, it is possible to provide
necessary and sufficient conditions for the formation of apparent
horizons which are not overwhelmed by the gauge.

In order to be really satisfied that we have control over the
gauge, we propose in this paper to place our prejudice in favor of
extrinsic time foliations to the test, i.e., by abandoning such
foliations entirely in favor of a foliation which is specified
completely by a condition on the intrinsic geometry of the leaves of the
foliation. We will explore the simplest choice: that the spatial
geometry be flat. It is well known that
such slices are provided  in the Schwarzschild spacetime
geometry by Lema\^\i tre coordinates (see e.g. \cite{LEM}).
Recently, these coordinates have been exploited to describe the canonical
reduction of the theory with a source consisting of a shell of
massive dust \cite{FLW}. Indeed,
there has been a revival of interest in the canonical reduction of
spherically symmetric general relativity (see, for example,
\cite{Kuc,Haj} and references therein). However,
when the sources are not confined to a shell,
the canonical procedure in an extrinsic
time foliation is not generally tractable. The notable exception
is the polar slicing \cite{BCMN,Unruh}.
But simplicity comes at a price: the corresponding
chart is pathological if the slice possesses an apparent horizon, as it
ultimately will in a classically collapsing geometry.
Flat slicings, while remarkably simple,
escape this shortcoming.

Historically, flat foliations of spacetime are of interest due to
the role they played in the seventies as a  provider of useful
guidelines for the construction of the proof of the positive mass
theorem. Brill and Jang \cite{BJ} showed that if the spatial geometry is
flat, and the extrinsic curvature falls off more rapidly than
$r^{-3/2}$ at infinity, the only
solution of the vacuum Einstein equations is flat.
One might, therefore, be tempted to conclude that
such foliations are pathological, inconsistent as they
appear to be with the
presence of gravitational waves. As we will see, however,
the falloff of the extrinsic curvature in this gauge is slow
(exactly as $r^{-3/2}$) and thus Brill and Jang's argument is not
applicable. Indeed, this slicing possesses the 
unusual feature that the ADM mass is encoded completely by the 
extrinsic curvature. While we do not have
gravitational waves to contend with in a spherically symmetric
geometry, it is still possible  that the gauge
exhibit pathologies associated with the sources.
However, we demonstrate here that if the sources are finite, all
solutions of the constraints are non-singular if they are non
singular at the origin.
Therefore, this gauge steers clear of singularities
because the extrinsic curvature (which is now the only
measure of the geometry) is always bounded when the
sources are. This should be contrasted with the situation
in an extrinsic time foliation where the
solutions may well be singular if the sources are large but
finite \cite{II,III}.

We examine the occurence of apparent horizons
in the intrinsic time gauge.
We must abandon the geometrically satisfying
but nonetheless gauge dependent identification of an
apparent horizon with the formation of a bag with a neck
(minimal surface) in the
spatial geometry. Because the geometry is flat, there is no bag to
speak of, much less the kind of bag envisaged by Wheeler.
Now apparent horizons occur
due to the action of extrinsic curvature alone.
However, it is no longer enough that the extrinsic curvature be
large, it also must possess the appropriate sign.
This is due to the peculiarity of this gauge that, when the
dominant energy condition is satisfied, the tangential projection
$K_R$ of the extrinsic curvature  tensor has a definite sign.

We next search for necessary and sufficient conditions for the
appearance of apparent horizons. Remarkably, once we take into
account the obvious obstruction associated with the
sign of $K_R$, we can reproduce
the inequalities we derived earlier in the extrinsic time gauges
\cite{IV,V}. This suggests a validity which transcends their gauge
dependent derivation.

It remains to confirm that the flat foliation is consistent with
the dynamics. We do this by constructing explicitly both the
lapse and the shift necessary to maintain the
flatness of the intrinsic geometry and we write down explicitly the
exterior form of the spacetime metric.

Schoen and Yau in \cite{SY} gave a sufficient 
condition for the appearance of
trapped surfaces in general initial data sets. It turns out that the only 
place these estimates can be realistically tested is on intrinsically flat
manifolds. We compare the Schoen and Yau sufficient condition with ours 
for the
case of a constant density intrinsically flat sphere and find that ours is
better by a factor of 3.

Finally, in the same spirit which has prompted us to consider flat
foliations in the first place, we examine the flat
foliation as a special case of foliations with negative spatial
scalar curvature.

\vskip 1pc
\noindent{\bf  Flat Foliations}
\vskip 0.5cm
We recall that the constraints are given by (we exploit the
notation introduced in \cite{I})

\begin{equation}
K_R\left[K_R+2K_{\cal L}\right]-
{1\over R^2}\Big[2 \left(R R' \right)' -R'^2 -1 \Big]=8\pi \rho\,,
\label{eq:2.1a}
\end{equation}
and

\begin{equation}
K_R' + {R' \over R}(K_R-K_{\cal L})=4\pi J\,, \label{eq:2.1b}
\end{equation}
where the line element of the spherically symmetric spatial
geometry is parametrized by

\begin{equation}
ds^2= d\ell^2+R^2 d\Omega^2\,,
\label{eq:2.1c}
\end{equation}
and we have expanded the extrinsic curvature ($n^a$ is the
outward pointing unit normal to the two-sphere of fixed $\ell$) as,

\begin{equation}
K_{ab}= n_a n_b K_{\cal L} + (g_{ab}-n_a n_b)K_R\,. \label{eq:2.1d}
\end{equation}
All derivatives are with respect to the proper radius of the
spherical geometry, $\ell$. The spatial geometries we wish to
consider are $R^3$ ($\ell\in [0\infty)$ with 
a single asymptotically flat region with a
regular center, $\ell=0$. The appropriate boundary condition on
the metric at $\ell=0$ is then

\begin{equation}
R(0)=0\,.
\label{eq:2.2}
\end{equation}
We recall that $R'(0)=1$ if the geometry is regular at this point.
We assume that both the energy density of matter, $\rho$, and its
radial flow, $J$, are appropriately bounded functions of $\ell$ on
some compact support.

When we consider general spherically symmetric initial data we
 start off with six functions, $(g_{rr}, g_{\theta \theta}, K_R,
K_{\cal L}, \rho, J)$, which satisfy the two constraints. Because
the dynamics reside in the matter field, so that the geometry
is purely kinematical, it seems natural to choose $(\rho, J)$ as
the independent variables. This still leaves us with four dependent
objects satisfying the two constraints. One of the extra degrees of
freedom is obviously the coordinate choice on the three slice. We
can fix this more-or-less independently of everything else. One
natural choice (which we use) is to set $g_{rr} \equiv 1$; the other standard
choices are to arrange that the metric be conformally flat or that
the radial coordinate be the areal (Schwarzschild) radius.

This leaves us with one extra variable among the three $(g_{\theta
\theta} \equiv R^2, K_R, K_{\cal L})$ so we choose some
relationship between them. Such a condition should fix the slicing,
the way that the given slice is embedded into the spacetime.
When we solve the constraints completely
we determine both the intrinsic geometry and the extrinsic
curvature.

 In this paper  we propose to foliate spacetime using the
intrinsic geometry to  mark time. The simplest possible choice is
the one we will adopt.  Let us suppose that

\begin{equation}
R(\ell) = \ell
\label{eq:flat}
\end{equation}
everywhere so that
the spatial geometry is flat $R^3$ everywhere. This condition is,
{\it a priori}, no more restrictive than any of the other slicing
conditions we have considered. The scalar  curvature now vanishes
with the result that the hamiltonian  constraint reduces to the
algebraic condition on
$K_{ab}$ in terms  of $\rho$,

\begin{equation}
K_R\left[K_R+2K_{\cal L}\right]
=8\pi \rho\,.
\label{eq:Constr2a}
\end{equation}
The momentum constraint, apparently at least, is only modified in
a trivial way: $R'/R$ is replaced by $1/\ell$.

To solve the constraints we eliminate the extrinsic curvature
scalar $K_{\cal L}$ from Eq.(\ref{eq:Constr2a}) in favor of $K_R$
and $\rho$ and substitute into (\ref{eq:2.1b}). We get

\begin{equation}
(\ell^{3/2} K_R)' = 4\pi \ell^{3/2}({\rho\over \ell K_R} + J) \,.
\label{eq:Constr2c}
\end{equation}
The equation is no longer linear in $K_R$. This is not too
surprising. The constraints in general relativity are non-linear to
start with and the physics is now completely encoded in the
extrinsic geometry.

\vskip 1pc
\noindent{\bf Lema\^\i tre Slicing of the Schwarzschild Geometry}
\vskip 1pc

As mentioned above, one example of such a flat slicing is given by
the Lema\^\i tre coordinatization of the Schwarzschild solution.
We start off with the Schwarzschild solution in standard
coordinates
\begin{equation}
ds^2 = -\left(1 - {2m \over R}\right)dt^2
+ \left(1 - {2m \over R}\right)^{-1}dR^2 + R^2 d\Omega^2\,,
\label{lem1}
\end{equation}
and change the time coordinate via
\begin{equation}
t = \tau + 4m\sqrt{{R \over 2m}} + 2m\ln \left|{\sqrt{R} -
\sqrt{2m}
\over \sqrt{R} + \sqrt{2m}}\right|\,. \label{eq:lem2}
\end{equation}
Differentiating Eq.(\ref{eq:lem2}), we get

\begin{equation}
dt = d\tau + \left(1 - {2m \over R}\right)^{-1}\sqrt{{2m \over
R}}dR\,,
\label{lem3}
\end{equation}
and when this is substituted into Eq.(\ref{lem1}) we obtain

\begin{equation}
ds^2 = -\left(1 - {2m \over R}\right)d\tau^2
- 2 \sqrt{{2m \over R}}dR d\tau + dR^2 + R^2 d\Omega^2\,.
\label{lem4}
\end{equation}
It is clear from Eq.(\ref{lem4}) that the spatial geometry of the
$\tau$ equal constant slices is flat \cite{CO}.
In adition, Lema\^\i tre coordinates are non-singular on the horizon at $R=2m$.
The price one pays, however,
is that the form of the spacetime metric is no longer
diagonal in $R$ and $\tau$.
Leima\^\i tre coordinates cover half of the maximal extension of the
Schwarzschild geometry. For our choice, this is the region above the
principal diagonal on the Kruskal diagram. Note that the spatial
geometry is regular at the the Schwarzschild singularity at $R=0$.
Each spatial slice intersects this singularity.

It is simple to construct the extrinsic curvature tensor.
The Wald (\cite{W}) definition of the extrinsic curvature gives

\begin{equation}
\partial_0 g_{ab} = 2NK_{ab} + N_{a;b} + N_{b;a}\,.
\label{lem7}
\end{equation}
In the coordinate choice given by Eq.(\ref{lem4}) we have
$\partial_0 g_{ab} = 0$, $N \equiv 1$, and $N_i = (-\sqrt{2m/R}, 0,
0)$ which gives

\begin{equation}
K_{RR} = -{1 \over 2}\sqrt{{2m \over R^3}}, \hskip 1cm
K_{\theta\theta} = +\sqrt{2mR}, \hskip 1cm K_{\phi\phi} =
+\sqrt{2mR}\sin^2\theta\,. \label{lem8}
\end{equation}
From the definition(\ref{eq:2.1d}) we get
\begin{equation}
K_{\cal L} = -{1 \over 2}\sqrt{{2m \over R^3}}, \hskip 1cm
K_R = +\sqrt{{2m \over R^3}}\,. \label{lem9}
\end{equation}
Clearly these satisfy Eqs.(\ref{eq:Constr2a}),
and (\ref{eq:Constr2c}) with vanishing sources.
Both $K_{\cal L}$ and $K_R$ diverge at $R=0$.
The singularity in the spacetime curvature is encoded completely in the
extrinsic curvature.

\vskip 1pc
\noindent{\bf `Momentarily static solution'}
\vskip 1pc

A peculiarity of the flat foliation is that
$K_R$ responds directly to $\rho$. Unlike its behavior in 
extrinsic time foliations,
$K_R$ does not vanish if $J$ is identically zero unless $\rho$ also
is identically zero. Clearly,
if there are any sources acting, this foliation does not admit any
moment of time symmetry solutions with $K_{ab}=0$.
In this respect, the simple constant density `star'
solutions, with $\rho$ constant on some compact support and
$J=0$, should not be confused with genuine constant density
solutions occurring in a momentarily static configuration.

\vskip 1pc
\noindent{\bf Two solutions for each $(\rho,J)$}
\vskip 1pc

One can show using Eq.(\ref{eq:Constr2c}) that there are two solutions
for each specification of $(\rho,J)$. We write Eq.(\ref{eq:Constr2c}) as
\begin{equation}
K_R' + {3 \over 2}\ell^{-1}K_R = 4\pi\ell^{-1}{\rho \over K_R} +
4\pi J~.\label{eq:0}
\end{equation}
Suppose that $K_R$ is finite at the origin, $\ell=0$. Then

\begin{equation}
K_R(0) = \pm\sqrt{8\pi\rho(0)\over 3}\,, \end{equation}
which is independent of $J$. Having fixed
$K_R(0)$, Eq.(\ref{eq:0}) allows us to integrate out to find
$K_R(\ell).$

\vskip1pc
\noindent{\bf The momentum constraint and the quasi-local mass}
\vskip1pc

An alternative casting of the momentum constraint is as

\begin{equation}
(\ell^3 K_R^2)' = 8\pi \ell^2(\rho + J \ell K_R ) \,.
\label{eq:Constr2d}
\end{equation}
We recall that in this gauge the spherical quasi - local mass $m$
(the Misner-Sharp, Hawking, {\it et al} mass) \cite{MS,HAWK}
(see also \cite{MOM} and \cite{I} for references), defined by

\begin{equation}
m(\ell) = {1\over 2} R(1 - R^{\prime2}) + R^3K_R^2\,,
\end{equation}
is completely determined by extrinsic curvature:

\begin{equation}
m(\ell) = {1 \over 2}\ell^3 K_R^2 \,,
\end{equation}
and our rewriting of the constraint, Eq.(\ref{eq:Constr2d}),
can be identified with the integrability condition
on $m$,

\begin{equation}
m' =
4\pi R^2(\rho R' + J R K_R ) \,.
\label{eq:m'}\end{equation}
Thus we
can solve Eq.(\ref{eq:Constr2d}) as
\begin{equation}
K_R = \pm {1\over \ell^{3/2}} \sqrt{2m(\ell)}\,,
\label{eq:m}
\end{equation}
where $m(\ell)$ is the quasi-local mass.
Therefore, a necessary condition for the
existence of a flat slice through a given spherical spacetime is
that the mass function be positive.
$m$ is manifestly positive in this gauge.

The geometry is necessarily singular if the  constant of
integration $m(0)$ is non vanishing. Indeed, in vacuum,
$m(\ell)=m(0)$  and we reproduce a Lema\^\i tre slice of the
Schwarzschild geometry.

Outside the support of matter,

\begin{equation}
K_R^2 = {2 m(\ell_0)\over \ell^{3}} \,.
\end{equation}
The ADM mass is encoded completely in $|K_R|$. We note also that
$K_R$ has the same asymptotic falloff as the gauge,
$K_{\cal L} + 0.5 K_R=0$.
In this gauge, the Hamiltonian constraint
assumes the intrinsic geometric form,

\begin{equation}
^3{\cal R} =16\pi\rho\,,
\label{eq:R}
\end{equation}
where $^3{\cal R}$ is the scalar curvature.
It is equally legitimate, however, to
treat Eq.(\ref{eq:R}) itself as a source dependent
intrinsic time gauge condition.  The
Hamiltonian constraint then reduces to the
algebraic condition, $K_{\cal L} + 0.5 K_R=0$. This way, the extrinsic
time gauge can be viewed as entirely equivalent to
an intrinsic time gauge, albeit an unconventional one depending explicitly
on the source.

\vskip1pc
\noindent{\bf $J=0$ is exactly solvable}
\vskip0.5pc

 When
$J=0$, for any given
$\rho$, Eq.(\ref{eq:Constr2d}) is exactly solvable. Suppose that
the weak energy condition,
$\rho\ge 0$, holds. Then

\begin{equation}
K_R = \pm {1\over \ell^{3/2}} \sqrt{2M(\ell)}\,,
\label{eq:J=0}
\end{equation}
where \cite{BIND}

\begin{equation}
M(\ell) = 4\pi \int_0^\ell d\ell\, \ell^2\, \rho \,,
\end{equation}
is the bare material mass and equals the mass function. It must be
positive. One solution is minus the other and $K_R$ has a definite
sign. Indeed, when the dominant energy condition holds, this property
continues to hold.

\vskip1pc
\noindent{\bf $K_R$ has a definite sign}
\vskip0.5pc

If
$\rho\ge c |J|$ for any constant $c$, then $K_R$
possess a definite sign. In particular, this is true if the
dominant energy condition, $c=1$,  is satisfied.
To prove this, let us suppose that $K_R$
is positive at $\ell=0$. Suppose that it falls to zero
on the support of $\rho$. Where $K_R\sim 0+$, the combination
${\rho/\ell K_R} + J$ appearing on the RHS of Eq.(\ref{eq:Constr2c})
is positive so that the LHS, $(\ell^{3/2} K_R)'$, is also positive.
$K_R$ is therefore increasing and cannot fall through zero.
An identical argument applies for negative $K_R$.
In this gauge, dominant energy places a very strong constraint on $K_R$.
Note, however, that the radial extrinsic curvature, $K_{\cal L}$,
which is  now determined by Eq.(\ref{eq:Constr2a}), does not possess a
definite sign.

The effect of introducing a current source, $J$, on a
solution with $J=0$ is easily deduced
from Eq.(\ref{eq:Constr2d}). If $K_R$ is positive, then a positive
current increases $K_R$, whereas a negative one decreases it. In particular,
if $J$ has a definite sign (positive say), the  solution
has everywhere greater $K_R$ than the corresponding solution with $J=0$.

\vskip 1pc
\noindent{\bf  The extrinsic geometry is non - singular everywhere}
\vskip1cm

When $\rho$ and $J$ are finite, and the center is non-singular,
$K_R$ is finite and bounded away from zero everywhere. However,
the only way the extrinsic geometry can become singular is by
$K_R$ diverging or $K_{\cal L}$ diverging (for
which it is necessary that $K_R=0$)
so that the flat slice is singularity avoiding.

We can place an explicit bound on $K_R$ as follows. We integrate
Eq.(\ref{eq:Constr2d})

\begin{equation}
L^3 K_R^2 = 8\pi \int_0^L d\ell\,
\ell^2(\rho + J \ell K_R ) \,.
\label{eq:ineq1}
\end{equation}
We then have

\begin{equation}
L^2  (K_R^2)_{\rm Max} \le 8\pi \left({1\over 3}
L^2\rho_{\rm Max}
+ {1\over 4} L^3 J_{\rm Max}
(K_R )_{\rm Max}\right)
\,, \label{eq:ineq20}
\end{equation}
so that (by solving the quadratic)

\begin{equation}
L(K_R )_{\rm Max} \le
 \pi J_{\rm Max}L^2
+ \left[\pi^2 \Big(J_{\rm Max}L^2\Big)^2
+ {8\pi\over 3} \,\rho_{\rm Max}L^2 \right]^{1/2}\,.
\label{eq:ineq2}
\end{equation}
$K_R$ is clearly finite when $\rho$ and $J$ are.
We note that the bound is more sensitively dependent on $J_{\rm
Max}$ than it is on $\rho_{\rm Max}$. Finally, we have seen that
$K_R$ cannot vanish on the support of $\rho$ when the (generalized)
dominant energy condition is satisfied.
Thus $K_{\cal L}$, which is  determined by Eq.(\ref{eq:Constr2a}),
is also bounded.

\vskip 1pc
\noindent{\bf Apparent Horizons}
\vskip 1pc

We define the two optical scalars,

\begin{equation}
\omega_\pm = 2(R' \pm R  K_R)\,.
\label{eq:ah}
\end{equation}
A future (past) apparent horizon forms when $\omega_\pm =0$. The
surface is future (past) trapped when $\omega_\pm \le 0$. In this
gauge, the appearance of an
apparent horizon is entirely due to the action of extrinsic
curvature.

We have already seen that solutions fall into two categories
when the dominant energy condition is satisfied: those with $K_R>0$ and
those with $K_R<0$.
On solutions with  $K_R$ positive (negative), future
(past) trapped surfaces are impossible.
Let us therefore focus on
the occurence of future (past) trapped surfaces in
solutions with $K_R <0$ ($K_R>0$).

Suppose that  $K_R<0$ and that the
solution is free of future apparent horizons so that
$\ell K_R > -1$. It is then clear from
Eq.(\ref{eq:Constr2c}) that $(\ell^{3/2} K_R)' <0$ so that $\ell^{3/2}K_R$
decreases monotonically from zero at the origin (it saturates at
the   value $-\sqrt{2m}$ outside the source).
Thus, if $\ell^{3/2}K_R$ is not monotonic, the solution must possess
an apparent horizon.
Of course the converse of this statement is false: monotonic
$\ell^{3/2}K_R$ does not necessarily imply that the
geometry is free of an apparent horizon (recall that any solution with $J=0$
given by Eq.(\ref{eq:J=0}) is monotonic).

A sufficient condition for the formation
of trapped surfaces can be obtained as follows:

We integrate Eq.(\ref{eq:Constr2c}) from $\ell=0$ up to $\ell=L$:

\begin{equation}
L^{1/2}(\omega_+ -1)_{\ell=L} =
4\pi \int_0^L d\ell\,\ell^{3/2}({\rho\over \ell K_R} + J) \,,
\label{eq:Suff1}
\end{equation}
Let us suppose that the surface
at $\ell=L$ is not future trapped, so that $\omega_\pm \ge 0$, nor does
any trapped surface exist in the interior ($\ell K_R \ge - 1$).
Then

\begin{equation}
L^{1/2} \ge
4\pi\int_0^L d\ell\, \ell^{3/2}\,({\rho\over -\ell K_R} - J) \,.
\label{eq:Suff2}
\end{equation}
If $\ell K_R > 0$, the inequality is vacuous as we would expect.
If $\ell K_R <0$, then $1/(-\ell K_R) > 1$, so that

\begin{equation}
L^{1/2} \ge
4\pi\int_0^L d\ell\, \ell^{3/2}\,(\rho - J) \,. \label{eq:Suff3}
\end{equation}
In addition,

\begin{equation}
\int_0^L \ell^2 f(\ell) d\ell \le L^{1/2}  \int_0^L \ell^{3/2} f(\ell)
d\ell\,,
\label{eq:power}\end{equation}
for any positive function, $f(\ell)$, so that, with dominant energy, we obtain

\begin{equation}
L \ge
4\pi\int_0^L d\ell\, \ell^2\,(\rho - J)\, = \, M - P
\,,
\label{eq:Suff4}
\end{equation}
where

\begin{equation}
P=4\pi \int_0^L d\ell\,\ell^2\, J\,.
\end{equation}
It is clear that Eq.(\ref{eq:power}) is sharp as it is saturated
with $f$ peaked sharply about $L$.

The inequality Eq.(\ref{eq:Suff4}) depends only
on the physical measures of the initial data, i.e., $\rho$, and
$J$, and the size of the region, $L$. It assumes the same form as
the inequality we obtained in \cite{IV}. Indeed,
 in this gauge, we equal the best constant we obtained in \cite{IV}.

Note that if the dominant
energy condition is violated, the inequality
Eq.(\ref{eq:power}) is not valid.
No inequality analogous to Eq.(\ref{eq:Suff4}) appears to hold.

We can exploit the universal bound on $K_R$, Eq.(\ref{eq:ineq2}) to obtain a
corresponding necessary condition.

Let the first apparent horizon occur at $\ell =L$. Then

\begin{equation}
L = 8\pi \int_0^L d\ell\,
\ell^2(\rho + J \ell K_R ) \,,
\label{eq:nec1}
\end{equation}
so that

\begin{equation}
L \le {8\pi L^3\over 3} \rho_{\rm Max}
+ 2\pi L^4 J_{\rm Max}(K_R )_{\rm Max}
\,. \label{eq:nec2}
\end{equation}
We now exploit the bound Eq.(\ref{eq:ineq2}) for
$L(K_R)_{\rm Max}$ to obtain

\begin{equation}
1 \le {8\pi \over 3} \rho_{\rm Max} L^2
+ {2\pi} J_{\rm Max} L^2\left[
 \pi J_{\rm Max}L^2
+ \left(\pi^2 (J_{\rm Max}L^2)^2
+ {8\pi\over 3} \,\rho_{\rm Max}L^2 \right)^{1/2}\right]\,.
\label{eq:1}\end{equation}
On rearrangement, Eq.(\ref{eq:1}) can be cast

\begin{equation}
1 \le 4\pi^2 \Big(J_{\rm Max}L^2\Big)^2 + {8\pi\over 3} \,\rho_{\rm
Max}L^2  \,.
\end{equation}
If, in addition, we exploit dominant energy, we can replace
$J_{\rm Max}$ by $\rho_{\rm Max}$ and solve the quadratic to get

\begin{equation}
{1\over 3\pi} \Big( {\sqrt{13}\over 2} -1\Big)
\le \, \rho_{\rm Max}L^2 \,.\label{eq:Suff5}
\end{equation}
Thus if $\rho_{\rm Max} L^2 < (\sqrt{13}/2 -1)/3\pi$,
the region cannot contain an apparent horizon.
The constant of proportionality is comparable to that appearing
in Eq.(57) of \cite{V} with $\alpha=1$.
The derivation is, however,  considerably simpler. In the
extrinsic time foliation, the derivation depended in an essential
way on the application of a weighted Poincar\'e Inequality.
This improvement is clearly related to the
singularity avoidance of the intrinsic time gauge we have exploited
here.

\vskip2pc
\noindent{\bf Consistency of flat foliation with
the evolution}
\vskip1pc
The condition that the flat foliation be preserved under
evolution,
$\partial_0 \gamma_{ab}=0$, implies that the extrinsic curvature is
proportional to a Killing form,

\begin{equation}
2 N K_{ab} = - \nabla_a  N_{b} - \nabla_b N_a \,.
\end{equation}
In the spherically symmetric geometry we are considering this reduces to
the set of conditions,

\begin{eqnarray}
N_\ell' &=& N K_{\cal L}\nonumber\\
N_\ell &=& \ell N K_R\,.
\end{eqnarray}
Here $N$ and $N_\ell$ are respectively the lapse and the radial shift.
Eliminating $N$, we obtain,

\begin{equation}
N_\ell' = {1\over \ell} {K_{\cal L}\over K_R} N_\ell\,,
\end{equation}
with solution,

\begin{eqnarray}
N_\ell &=& \exp( - \int d\ell \alpha /\ell )\nonumber\\
N &=&  \exp( - \int d\ell \alpha /\ell) / \ell K_R\,,
\label{eq:N}
\end{eqnarray}
where $\alpha = - K_{\cal L}/K_R$ is the ratio of extrinsic curvature
scalars introduced in \cite{III}.
Thus flat slicing is consistent with evolution. Indeed, both the
lapse and shift are completely determined without appealing to the
dynamical Einstein equations for $K_{\cal L}$
and $K_R$.

In the exterior region we have
$\alpha = 1/2$, so that
$N_\ell = N_\ell(L) (L/\ell)^{1/2}$ and
$N = N_\ell(L) L^{1/2} (2m_0)^{-1/2}$ where $N_\ell(L)$
is the boundary shift. The lapse is constant.
The boundary condition $N\to 1$ at infinity therefore fixes $N_\ell(L)$.
Let $\tau$ be the time coordinate defined by this foliation.
We then have the exterior spacetime metric

\begin{equation}
ds^2 =
- \left(1 -{ 2m_0\over \ell}\right) d\tau^2 \pm 2\sqrt{{2m_0\over
\ell}} d\tau d\ell + d\ell^2 + \ell^2 d\Omega^2\,.
\end{equation}
This is the Schwarzschild metric expressed in Lema\^\i  tre
coordinates as we have already seen in Eq.(\ref{lem4}). The
spacetime is completely characterized by the shift.

\vskip2pc
\noindent{\bf Negative scalar curvature}
\vskip1pc

Let us consider instead of the flat slicing, any
foliation with negative scalar curvature,

\begin{equation}
^3 {\cal R} = -
{1\over R^2}\Big[2 \left(R R' \right)' -R'^2 -1 \Big]\,.
\label{eq:curv}
\end{equation}
 It is then simple to demonstrate that
$R'\ge 1$ everywhere so that $R \ge \ell$.
We have, at a critical point of $R'$,

\begin{equation}
R'^2 = 1-
{}^3{\cal R} \,\,R^2 \,.
\label{eq:curv1}
\end{equation}
If the scalar curvature is bounded then so is $R'$.
Such foliations
are clearly very different from the extrinsic time foliations we
considered in \cite{III,IV,V} with everywhere positive scalar curvature
in which $R'^2\le 1$ and  $R\le \ell$ in regular solutions.
With a  prescribed value of the
scalar curvature, we can solve the constraints exactly as we
did in the flat slice. We get,

\begin{equation}
(R^{3/2} K_R)' = 4\pi R^{3/2}({R'\tilde\rho\over R K_R} + J) \,,
\label{eq:Constr2f}
\end{equation}
where we set $\tilde \rho = \rho - ^3{\cal R}/16\pi$. We have
$\tilde\rho\ge \rho$ so that whatever energy condition is
good with  $\rho$ is better with $\tilde\rho$.
Now, exactly as in a flat slicing, $K_R$ has a definite sign when
dominant energy holds.

It is straightforward to demonstrate that a bound can be placed on
$K_R$. Such gauges therefore share with the flat foliation its
singularity avoidance. This differs from any foliation with $^3{\cal R}$
positive where large sources are not always consistent
with any singular geometry.

Let us examine the robustness of Eq.(\ref{eq:Suff4}).
It is simple to demonstrate that instead of
Eq.(\ref{eq:Suff4}), we obtain the sharp inequality,

\begin{equation}
R R'  \ge   \, M -  P
\,,
\label{eq:Suff44}
\end{equation}
which does not depend explicitly on ${}^3{\cal R}$.
Eq.(\ref{eq:Suff44}) coincides with Eq.(\ref{eq:Suff4}) when $R=\ell$.
We see, however that the more negative the  scalar curvature,
the greater is the maximum of  $R'$, and thus the weaker the
inequality.

\vskip2pc
\noindent{\bf The Schoen and Yau criterion for trapped surfaces}
\vskip1pc

The first mathematically precise statement of a sufficient condition for the
appearance of apparent horizons in a general (i.e. nonsymmetric) initial data
set was given by Schoen and Yau in \cite{SY}. This condition is difficult to
evaluate in general but it is possible to use it in the special case where 
the spatial geometry is flat, exactly the situation we are discussing in this
article.

Schoen and Yau define the size of any compact three dimensional subset $(\Omega
)$ of a reimannian manifold as the minor radius of the largest three torus that
can be embedded in $\Omega$. They call this $Rad(\Omega)$. In \cite{SY} they
prove two theorems. Theorem I is a statement that one cannot have a large set
with large positive scalar curvature. More precisely, they show that if the
scalar curvature of $\Omega$ is bounded below by a positive constant, 
$^{(3)}{\cal R}
\ge {\cal R}_0 > 0$, then
\begin{equation}
Rad(\Omega)  \le   \, \sqrt{8\pi^2 \over 3{\cal R}_0}
\,.
\label{eq:SY1}
\end{equation}
Theorem II deals with the situation where one has a solution to the
initial value constraints (including positive matter) on a set $\Omega$. If the
matter satisfies $\rho - |J| \ge \lambda > 0$ and if $\Omega$ is large in the
sense that
\begin{equation}
Rad(\Omega)  \ge   \, \sqrt{3\pi^2 \over 16\pi \lambda}
\,,
\label{eq:SY2}
\end{equation}
then the initial data must have a trapped surface.

If we have a maximal slice, we have that $^{(3)}{\cal R}
 = K^{ij}K_{ij} + 16\pi \rho
\ge 16\pi \lambda$ and so we can use $16\pi \lambda$ in place of 
${\cal R}_0$ in
Theorem I. However, since $8/3 < 3$, we have from Theorem I that we can
never get maximal initial data to satisfy Eq.(\ref{eq:SY2}).
It is clear that the estimates leading to Theorem II are not sharp and some
number smaller than 3 would almost certainly suffice. Unfortunately, the
constant in Theorem I is also not sharp (see \cite{OM}) and the two constants
are linked. Therefore our only hope of finding a nontrivial system in which to
use Theorem II of Schoen and Yau is to look at nonmaximal initial data. In this
case there is at least the possibility of having large $\rho$, so as to satisfy
the condition in Theorem II, while simultaneously having small
$^{(3)}{\cal R}$ 
to escape the barrier that Theorem I imposes. When one looks at the
way that Theorems I and II are derived, to try and find a configuration that
satisfies the condition in Theorem II, it is clear that one wants to have no
current, because it works against $\rho$ and to have the mass density as
uniform as possible. Further, one wants as little TT part in the extrinsic
curvature as possible as that adds to the scalar curvature. Finally, the metric
should be as simple as possible. This leads one to consider the situation where
the intrinsic metric is flat, and thus eliminating any barrier due to
Theorem I;
the extrinsic curvature is pure trace; and the trace is constant, so that there
is no current and one has a constant mass density. The other great advantage of
the flat metric is that one can easily evaluate their `torus' measure of the
size of a set.

In our notation, this is equivalent to choosing
$K_R = K_{\cal L} = {1 \over 3}{\rm tr}\, K = $ 
constant. From Eq.(\ref{eq:2.1b}) this
gives $J = 0$. From Eq.(\ref{eq:Constr2a}) we get that the mass density is
constant and satisfies
\begin{equation}
\rho = \rho_0 = {3K_R^2 \over 8\pi}.\label{eq:SY3}
\end{equation}
If we have a spherical set of radius $L$ satisfying this solution (one can
think of it as part of a flat cosmology), it is clear from Eq.(\ref{eq:ah})
that the horizon appears when $|LK_R| =
 1$. The sufficient condition we have derived (Eq.(\ref{eq:Suff4})) when
applied to this special case gives $|LK_R| \ge \sqrt{2}$, and the necessary
condition (Eq.(\ref{eq:Suff5})) gives $|LK_R| \ge 0.84$. Happily, these numbers
lie on each side of 1!

If we apply the Schoen and Yau condition to an intrinsically flat constant
density sphere of radius $L$, we get that $Rad(\Omega) = L/2$ and the 
Schoen and Yau sufficient condition (Eq.(\ref{eq:SY2})) 
becomes $|LK_R| \ge \sqrt{2}
\pi$. This calculation shows that the Schoen and Yau Theorem II 
is not vacuous. However, their set is 4 times larger than
is required and their sufficiency condition 
a factor of 3 weaker than ours.

\vskip2pc
\noindent{\bf Conclusions}
\vskip1pc

We have examined the constraints in spherically symmetric
general relativity using an intrinsic time to foliate spacetime.
Specifically we have foliated spacetime with flat spatial slices.
The presentation of the initial data on a flat spatial
hypersurface is very different from that on a
hypersurface belonging to an extrinsic time foliation of
spacetime. The Hamiltonian constraint becomes an
algebraic constraint on the extrinsic curvature:
when the weak energy condition is satisfied,
trajectories in superspace lie completely inside the
superspace light-cone --- the complement in superspace
of the allowed region consistent with any standard
extrinsic curvature foliation ($K_R=0$ or ${\rm Tr}\, K =0$ or,
more generally, any of those considered in \cite{III}).

In this foliation, solutions of the constraints {\it do} exhibit
peculiarities: when the sources are
finite, there are no singular geometries satisfying the
constraints other than
those which contain a singularity at their center;
when the dominant energy
condition is satisfied, $K_R$ possesses a definite sign;
they do not admit minimal surfaces.
Despite this, we find that  the physical description they provide
of apparent horizons is completely consistent
with that in an extrinsic time foliation.
Not only do the natural measures of the material content for
necessary and sufficient conditions ($\rho_{\rm Max}$ and $M$ respectively)
coincide with those we found when we considered extrinsic time slices
but, in addition, the inequalities assume identical forms.

Analogous gauges are applicable
with other topologies. For example, in a closed cosmology with $S^3$
topology one could choose $R(\ell) = {\ell_0\over 2\pi}\sin (\pi\ell/\ell_0)$,
where $\ell_0$ is the interpolar distance.

It would be interesting to examine the
canonical reduction and subsequent quantization
of spherically symmetric general
relativity in this gauge. The fact that many of the features of extrinsic
time foliations which are problematic do not occur suggests that
flat foliations could provide a valuable alternative, in particular,
for the description of gravitational collapse.

Finally, there appears to be no immediate obstruction to the
construction of a foliation of a general asymptotically flat spacetime by
a gauge of the form, $^3{\cal R} =0$.


\vskip2pc
\noindent{\bf Acknowledgements}

\vspace{.3cm}

We gratefully acknowledge support from CONACyT grant no. 2110855-0118PE to JG
and Forbairt Grant SC/96/750 to N\'OM.


\begin{thebibliography}{99}
\bibitem{CO} A further coordinate transformation to a
freely falling coordinate
system can be made by setting \cite{LEM}
\begin{equation}
R = \left[{3 \over 2}\left( {\varrho \over \sqrt{2m}} +
\sqrt{2m}\tau
\right)\right]^{{2 \over 3}}\,.
\label{lem5}
\end{equation}
Using $\varrho$ as a spatial coordinate the Schwarzschild metric
can be written as
\begin{equation}
ds^2 = -d\tau^2 + \left[{3 \over 2}\left( {\varrho \over \sqrt{2m}}
+ \sqrt{2m}\tau
\right)\right]^{-{2 \over 3}}d\varrho^2
+ \left[{3 \over 2}\left( {\varrho \over \sqrt{2m}} + \sqrt{2m}\tau
\right)\right]^{{4 \over 3}}d\Omega^2\,.
\label{lem6}
\end{equation}
\bibitem{BIND} We have seen that when $J=0$ the quasilocal mass coincides
with the bare mass, $m = M$. In an extrinsic time
foliation of spacetime, the difference $m-M$
has been shown to be strictly negative and thus provides
a good measure of the binding energy of the spherically symmetric
system \cite{I}.
When $J\ne 0$ on a flat slice, however,
$m-M$ does not possess a definite sign because, from
Eq.(\ref{eq:m'}), we get that $(m - M)' = 4\pi J \ell^3 K_R$, which
can be either positive or negative. We conclude that
$m-M$ is not a satisfactory measure of  binding energy in this
gauge. A challenge
is to define a  quantity with a wider validity.
\end{thebibliography}
\end{document}